\documentclass[conference]{IEEEtran}
\usepackage{color,soul}
\IEEEoverridecommandlockouts
\usepackage{cite}
\usepackage{amsmath,amssymb,amsfonts}
\usepackage{algorithmic}
\usepackage{graphicx}
\usepackage{textcomp}
\usepackage{xcolor}
\usepackage{lipsum}

\usepackage{multirow}

\def\BibTeX{{\rm B\kern-.05em{\sc i\kern-.025em b}\kern-.08em
    T\kern-.1667em\lower.7ex\hbox{E}\kern-.125emX}}

\begin{document}

\title{{Performance Analysis of Millimeter Wave Radar Waveforms for Integrated Sensing and Communication}}

\author{\IEEEauthorblockN{Akanksha Sneh}
\IEEEauthorblockA{\textit{Department of ECE} \\
\textit{Indraprastha Institute of Information Technology}\\
Delhi, India\\
akankshas@iiitd.ac.in}
\and
\IEEEauthorblockN{Aakanksha Tewari}
\IEEEauthorblockA{\textit{Department of ECE} \\
\textit{Indraprastha Institute of Information Technology}\\
Delhi, India\\
aakankshat@iiitd.ac.in}
\and
\IEEEauthorblockN{Shobha Sundar Ram}
\hspace{2cm}\IEEEauthorblockA{\textit{Department of ECE} \\
\textit{Indraprastha Institute of Information Technology}\\
Delhi, India\\
shobha@iiitd.ac.in}
\and
\IEEEauthorblockN{Sumit J Darak}
\IEEEauthorblockA{\textit{Department of ECE} \\
\textit{Indraprastha Institute of Information Technology}\\
Delhi, India\\
sumit@iiitd.ac.in}

}

\maketitle

\begin{abstract} Next-generation intelligent transportation systems require both sensing and communication between road users. However, deploying separate radars and communication devices involves the allocation of individual frequency bands and hardware platforms.  
Integrated sensing and communication (ISAC) offers a robust solution to the challenges of spectral congestion by utilizing a shared waveform, hardware, and spectrum for both localization of mobile users and communication. Various waveforms, including phase-modulated continuous waves (PMCW) and frequency-modulated continuous waves (FMCW), have been explored for target localization using traditional radar. On the other hand, new protocols such as the IEEE 802.11ad have been proposed to support wideband communication between vehicles. This paper compares both traditional radar and communication candidate waveforms for ISAC to detect single-point and extended targets. We show that the response of FMCW to mobile targets is poorer than that of PMCW. However, the IEEE 802.11ad radar outperforms PMCW radar and FMCW radar. Additionally, the radar signal processing algorithms are implemented on Zynq system-on-chip through hardware-software co-design and fixed-point analysis to evaluate their computational complexity in real-world implementations.

\end{abstract}

%

\section{Introduction}

Next-generation intelligent transportation systems have identified the requirement of both radar-based sensing and communications to improve road safety and reduce congestion. Radars offer complementary benefits to other automotive sensors, such as cameras and lidars. Specifically, radars help detect targets 24x7 and under all weather conditions with fine Doppler velocity resolution. Currently, 77 GHz millimeter wave bands have been allocated for automotive radar \cite{pandey2020database}. This band supports wide bandwidths that facilitate fine range resolution of targets and miniaturized circuits that can be easily mounted on a vehicle. Concurrent with the research and development of automotive radars, significant research has focused on developing vehicle-to-everything communication for sharing three-dimensional environmental information. Existing vehicular communication protocols operate at sub-6 GHz frequencies and hence cannot support very high data rates (of the order of Gbps) at ultra-low latencies such as device-to-device based V2X \cite{asadi2014survey}, dedicated short-range communications \cite{kenney2011dedicated}, and cellular V2X \cite{wang2018cellular}. Therefore, shifting to millimeter-wave communications supporting wide bandwidths and significant data rates has been a research focus. The large-scale deployment of automotive radar and communication networks, however, poses the challenge of spectral congestion and hardware and software costs of managing two separate systems free of interference. Hence, researchers have proposed using ISAC to combine radar and communication functionalities on a common hardware with a shared spectrum. This approach provides a low-cost solution with the additional benefits of effectively managing interference and synchronization issues between both systems.

ISAC has been extensively explored in prior art with various waveforms. There have broadly been two approaches. The first is the radar-centric ISAC, where communication data is embedded with conventional radar waveforms. For example, PMCW for ISAC is explored in \cite{dokhanchi2019mmwave}, where communication symbols are embedded in the PMCW radar signal. Further, FMCW, commonly used in commercial automotive radars, has investigated communication symbols in an integrated waveform \cite{zhang2017waveform}. The second approach is communication-centric, where the communication protocol is used for both sensing and communication. Recently, several works have discussed the ISAC implementations upon the IEEE 802.11ad protocol \cite{kumari2015investigating, kumari2018ieee,duggal2020doppler} and demonstrate the target detection with fine range resolution along with lower sidelobes. In all of the above ISAC frameworks, sensing is done through radar signal processing (RSP) to localize the mobile users (MUs)/ targets. More recently, RSP algorithms have been implemented on edge platforms such as field programmable gate array (FPGA) and heterogeneous system-on-chip (SoC) \cite{zhong2016design}. These platforms support hardware flexibility and in-field upgradability, making them suitable for real-world deployment of RSP. 

In this paper, we compare the performance of different ISAC waveforms, including PMCW, FMCW, and IEEE 802.11ad protocol, for localizing the target in terms of range and Doppler velocity. Further, the RSP is realized on an edge platform, Zynq system-on-chip (SoC), comprising a multi-core ARM processor and field programmable gate array (FPGA) through hardware-software co-design (HSCD).

The paper is organized in the following manner. Firstly, we present signal modeling and processing of ISAC waveforms in Section~\ref{sec:signalmodel} followed by simulation setup in  Section~\ref{sec:simsetup}. Further, we present the simulation results along with hardware complexity analysis in Section~\ref{sec:Results} followed by a brief summary of our work in Section~\ref{sec:Conclusion}.

\vspace{-1mm}
\section{Signal Model and Processing}
\label{sec:signalmodel}
In this section, we discuss the radar and communication waveform model transmitted at the access point (AP) and the signal reflected back from the target and received at the AP receiver. Further, we discuss the steps to localize the target in terms of range and Doppler velocity.\\
\textbf{Transmitted Signal:} We consider four different candidate ISAC waveforms for the transmit signal. First is the FMCW, in which the frequency is modulated for each pulse repetition interval (PRI) according to the chirp factor. The next waveform is the PMCW that has been modeled similarly to what is given in \cite{dokhanchi2019mmwave}. Here, the phase of the waveform is modulated through a differential phase shift keying technique. Finally, we consider a radar waveform embedded with a Golay sequence code. This is modeled after the IEEE 802.11ad radar discussed in \cite{kumari2015investigating} where the channel estimation (CE) field of the preamble of the IEEE 802.11ad PHY packet is jointly used for both radar and communication purposes. Each packet constitutes a PRI, and the radar waveform is the 512-bit Golay sequence in CE, which is composed of a Golay complementary pair. Their autocorrelation functions exhibit sidelobes with equal magnitudes but opposite signs. However, The perfect autocorrelation property of the Golay sequence is distorted for a mobile target due to the difference in the Doppler phase shift for a moving target with a Doppler shift that results in a phase shift between the cross-correlation output of the two consecutive packets.  The above-mentioned limitation of the Golay sequence in standard 802.11ad PHY packet is overcome by exploiting the Doppler-resilient Golay sequences, which have been proposed in 
\cite{pezeshki2008doppler} using the Prouhet-Thue-Morse sequence \cite{allouche1999ubiquitous}. 

Let us denote the discrete transmitted signal as $s_{tx}[q]$. Each waveform comprises $Q$ samples within one PRI denoted as  $T_{pri}$, and $P$ PRIs within one coherent processing interval (CPI). Now, these discrete-time signals are then converted into analog signals at the AP as follows:
\par\noindent\small
\begin{align}
\label{eq:dac}
    \mathbf{s_{tx}}(t) =\sum_{p=0}^{P-1}\sum_{q=0}^{Q-1}\mathbf{s}_{tx}[qT_s]\delta\left(t-qT_s - pT_{pri}\right).
\end{align}\normalsize
 Subsequently, the signal undergoes amplification with energy $A_s$, convolution with a shaping filter $\mathbf{m}_T$, and is finally upconverted to the mmW carrier frequency $f_c$ as shown
\par\noindent\small
\begin{align}
\label{eq:upcoverted}
  \mathbf{s_{tx_{up}}}(t) = \sqrt{A_s}\left(\mathbf{s_{tx}}(t) \ast \mathbf{m_T}(t)\right)e^{+j 2\pi f_c t}.
\end{align} \normalsize 


\textbf{Received Signal:} The upconverted transmitted signal is scattered by the mobile target and is received at the AP receiver. Let us assume $B$ point scatterers in the environment, then the received signal for $p^{th}$ packet is given by:

\par\noindent\small
\begin{align}
\begin{split}
\label{eq:rx_sig}
 \mathbf{s_{rx}}_p[q] = \sum_{b=1}^B \sigma_b \mathbf{s_{tx}}\left[q-q_b \right]e^{-j2\pi f_{D_b}pT_{pri}} 
 +  \mathbf{\zeta}.
 \end{split}
\end{align} \normalsize
Here, $\sigma_b$ represents the strength of the $b^{th}$ target that captures the radar cross-section of the target and the two-way propagation path loss factor. Further, $q_b$ corresponds to the sample index according to the delay ($\frac{2r_b}{c}$), where $r_b$ denotes the one-way range of the target with respect to AP and $c$ represents the speed of light. Doppler shift $f_{D_b}$ relates t the Doppler velocity of the target ($v_b$) as $f_{D_b} = \frac{2v_b}{\lambda}$, where $\lambda$ denotes the wavelength for mmW frequency $f_c$. $\zeta$ is the additive white Gaussian noise at the AP receiver. 


\textbf{Radar Signal Processing:} The reflected echoes from the target are collected for all $P$ packets and arranged in the form of a radar rectangle ${\mathbf{S}}$ which is given as input to the RSP block comprising $Q$ fast time samples across rows and $P$ slow time samples across columns.
Firstly,  we perform the matched filtering in the frequency domain through the multiplication operation since time-domain matched filtering through correlation is computationally complex. We first convert the time-domain received signal for each $p^{th}$ packet to the frequency domain through fast Fourier transform: $\tilde{\mathbf{S}}[q_f,p] = \text{1D-FFT}\{{\mathbf{S}}[q,p], q = 1 \cdots Q\}$. We jointly estimate the range and Doppler by performing 1D-FFT across the $P$ packets in conjunction with matched filtering as  
\par\noindent\small
\begin{align}
\label{eq:MF_FFT}
\tilde{\mathbf{\chi}}[f_j] = \mathbf{diag}\left[\tilde{\mathbf{S}}\mathbf{w}_{f_j}\tilde{\mathbf{s}}^\dag \right].
\end{align} \normalsize
Here,  $\mathbf{w}_{f_j}  = [1\; e^{j \frac{2\pi}{\lambda} f_j T{pri}} \cdots e^{j \frac{2\pi}{\lambda} f_j (P-1)T_{pri}}]^T$ is the delay vector to implement 1D-FFT for every Doppler shift $f_j$, $j = 1 \cdots J$ and $\tilde{\mathbf{s}}$ is the frequency domain signal of $\mathbf{s_{tx}}$.  The Doppler search space is spanned from $-f_{D_{max}}$ to $f_{D_{max}}$ through $\Delta f$ increments, where, $f_{D_{max}}$ corresponds $\frac{1}{2PRI}$. In \eqref{eq:MF_FFT}, diagonal elements are being extracted out of the complex $Q \times Q$ matrix through $\mathbf{diag}$ operation. Finally, the output is obtained by performing inverse fast Fourier transform (IFFT) on the frequency domain matched filter output, as ${{\chi}}[r_q,f_j] = IFFT\{\tilde{{\chi}}[q,f_j]\}$, where $r_q$ denotes the range samples which is directly related to the $q^{th}$ fast time sample. The peak is obtained from $\mathbf{\chi}$, which localizes the MU in terms of range and Doppler.


\section{Simulation Setup}
\label{sec:simsetup}

\begin{figure}[htbp]
    \centering
    \includegraphics[scale=0.55]{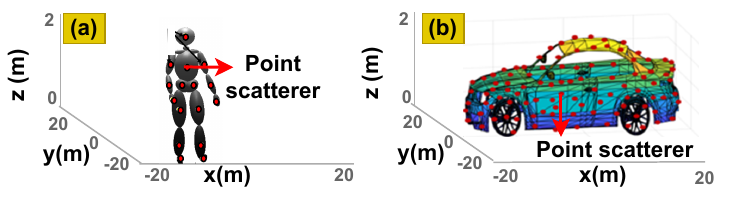}
    \caption{ \footnotesize Electromagnetic model of (a) pedestrian and (b) mid-size car.}
    \label{fig:target model}
\end{figure}

 We consider AP located at the Cartesian position coordinate of $[0,0,0]$ with the parameters as specified in Table~\ref{tab:RadarParam_Sim}. 
\begin{table}[htbp]
\scriptsize
    \centering
    \caption{\footnotesize Simulated Radar Parameters}
    \begin{tabular}{p{5cm}|p{2cm}}
    \hline \hline
     Parameters & Values  \\
    \hline \hline
        Centre frequency ($F_c$) & 60 GHz  \\ 
        Bandwidth ($BW$) & 1.76 GHz  \\
        Pulse repetition interval ($T_{PRI}$) & 2 $\mu$s \\
        Coherent processing interval ($T_{CPI}$) & 4 ms \\
          Maximum unambiguous range  & 44 m\\
        Maximum unambiguous velocity (m/s) & 625 m/s\\
        Range resolution & 0.085 m \\
        Velocity resolution & 0.3 m/s\\
      
    \hline \hline
    \end{tabular}
    \label{tab:RadarParam_Sim}
    \vspace{-2mm}
\end{table}
We consider two different types of electromagnetic targets which are commonly observed on roads. Both targets are extended targets modeled with multiple point scatterers that correspond to various parts of the targets' bodies. The first target is a pedestrian modeled with 27-point scatterers corresponding to ellipsoidal body parts as shown in Fig.\ref{fig:target model}(a). The data for the animated pedestrian in a walking motion are obtained from motion capture technology with the average RCS of 0 dBsm \cite{ram2008simulation}. The pedestrian is walking with a constant speed of $2$ m/s along a straight-line trajectory.  The second target is a car of dimensions $4.4m  \times 1.7m$ obtained from a freely available computer-aided design model, rendered with 6905 metallic plates, as shown in Fig.\ref{fig:target model}(b). The car is animated using the techniques explained in \cite{pandey2022classification}. The average RCS of the car is 10 dBsm. The car is assumed to be moving with a constant velocity of $10$ m/s along a straight-line trajectory. The targets are in direct line-of-sight with respect to the radar, and we ignore the effects of multipath shadowing.

\section{Performance Analysis}
\label{sec:Results}
\subsection{Simulation Performance Analysis}
\begin{figure}[htbp]
    \centering
    \includegraphics[scale=0.3]{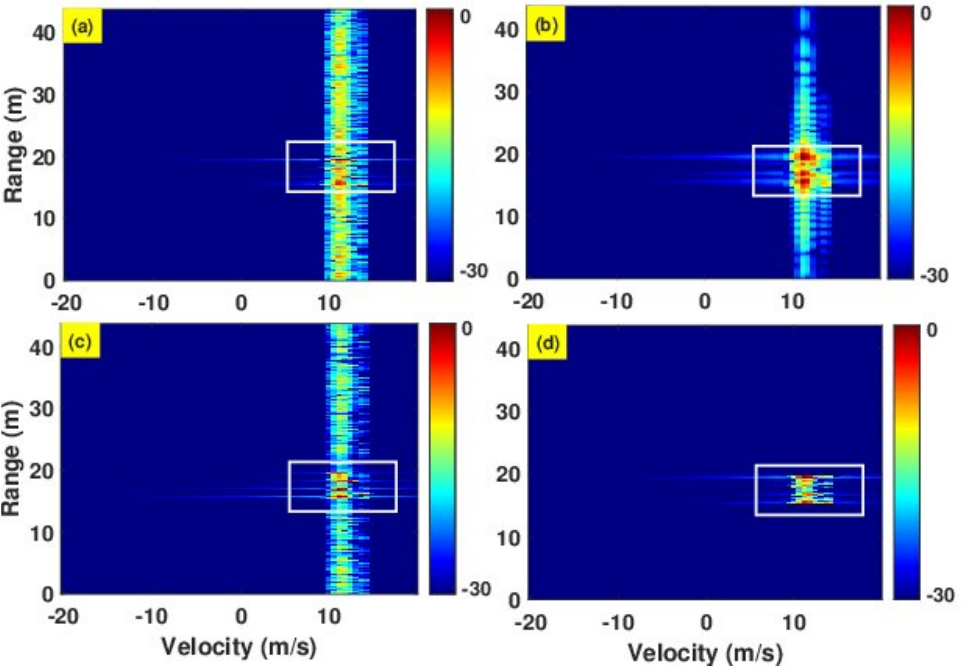}
    \caption{\footnotesize Localization of a mid-size car in the range-Doppler plane for (a) PMCW, (b) FMCW, (c) standard 802.11ad Golay sequence, and (d) Doppler-resilient 802.11ad Golay sequence waveforms.}
    \label{fig:range_doppler_plot_car}
\end{figure}

We analyze the performance of ISAC with the following waveforms: PMCW, FMCW, and conventional and Doppler-resilient Golay sequences present in the preamble of IEEE 802.11ad PHY frame structure. All the waveforms have identical sampling frequencies, pulse repetition intervals, and coherent processing intervals. Therefore, the range resolution, maximum unambiguous range, Doppler velocity resolution, and unambiguous Doppler are identical for a fair comparison. 
\begin{figure}[htbp]
    \centering
    \includegraphics[scale=0.03]{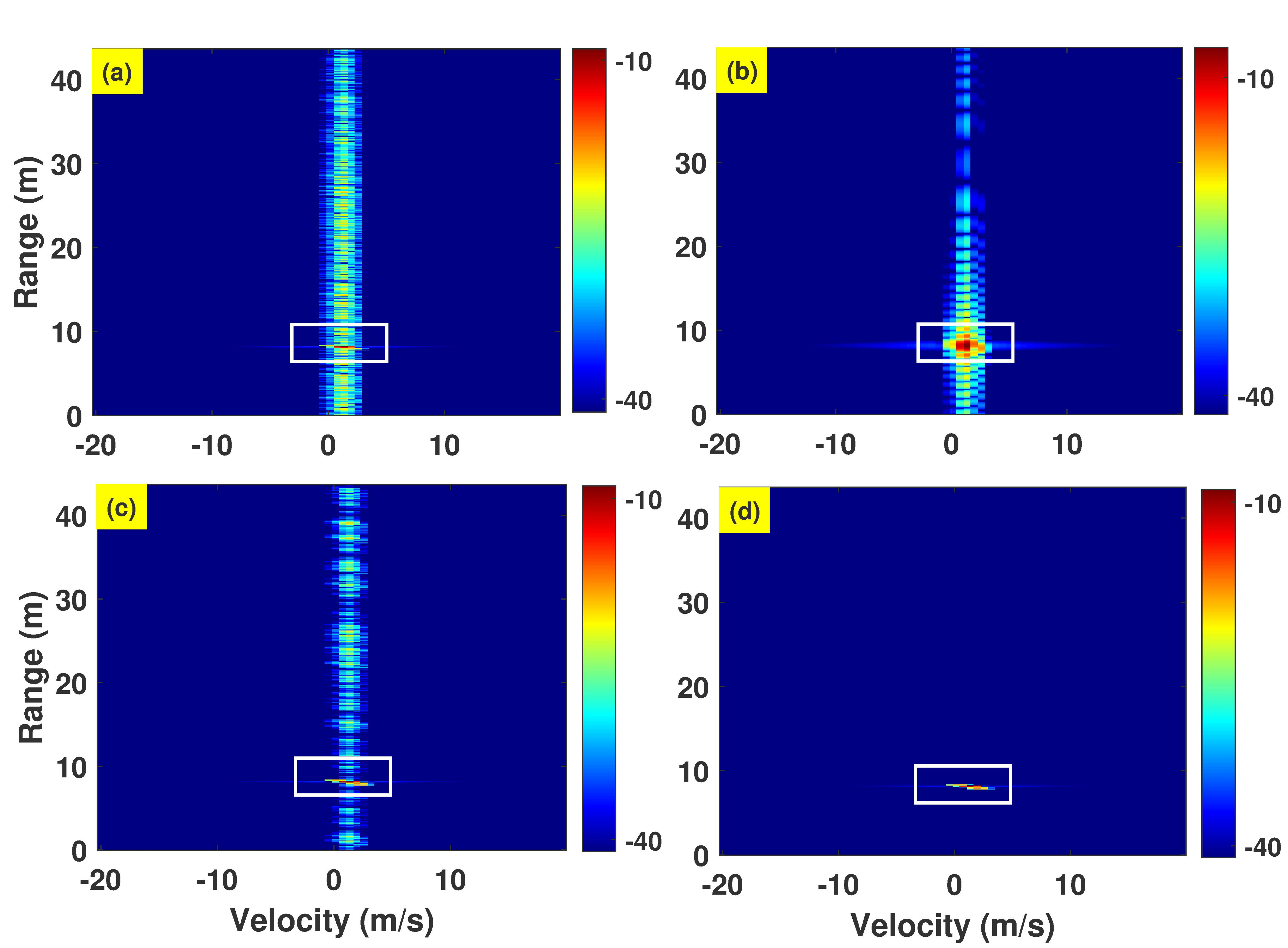}
    \caption{\footnotesize Localization of pedestrian in the range-Doppler plane for (a) PMCW, (b) FMCW, (c) standard 802.11ad Golay sequence, and (d) Doppler-resilient 802.11ad Golay sequence waveforms.}
    \label{fig:range_doppler_plot_human}
\end{figure}
First, we present the results regarding the localization of a pedestrian. The resultant range-Doppler (RD) ambiguity plots are presented for different waveforms in Fig.\ref{fig:range_doppler_plot_human}. For PMCW, Fig.\ref{fig:range_doppler_plot_human}(a) shows the spatial extent of the pedestrian along the range. The presence of the range sidelobes due to the motion of the pedestrian causes severe deterioration in the unambiguous RD plane. The high sidelobes make it harder to discern the exact spatial extent of the target. This degradation worsens for FMCW as shown in Fig.\ref{fig:range_doppler_plot_human}(b). The problem is significantly mitigated using the standard Golay sequences as shown in Fig.\ref{fig:range_doppler_plot_human}(c). However, we still observe the sidelobes of lower strength across the range dimension. This arises because the standard Golay sequence loses the hold of perfect autocorrelation property in the case of a moving target. Then we present the RD ambiguity plot for the Doppler-resilient Golay sequence in Fig.\ref{fig:range_doppler_plot_human}(d), where we observe almost zero sidelobes across the range for the moving target. Next, we present the RD ambiguity plots for a mid-size car as shown in Fig.\ref{fig:range_doppler_plot_car}. Here, we observe a wider target strength across the range as the target is spatially larger. In the case of PMCW, as shown in Fig.\ref{fig:range_doppler_plot_car}(a),  the high sidelobes degrade the target detection in the RD ambiguity plot. The performance has further deteriorated for FMCW as shown in Fig.\ref{fig:range_doppler_plot_car}(b) due to greater sidelobes. The results are slightly improved for the standard Golay sequence as shown in Fig.\ref{fig:range_doppler_plot_car}(c). However, the sidelobes still exist as the car moves with a constant velocity, causing a distortion in the autocorrelation property of the Golay sequences. Lastly, we analyzed the performance of the Doppler-resilient Golay sequence in Fig.\ref{fig:range_doppler_plot_car}(d)
where we observe an improved performance due to zero sidelobes, and hence, the exact spatial extent of the target is clearly visible.

\subsection{Hardware-Software Co-design and Fixed Point Analysis}

\begin{figure}
    \centering
    \includegraphics[scale=0.4]{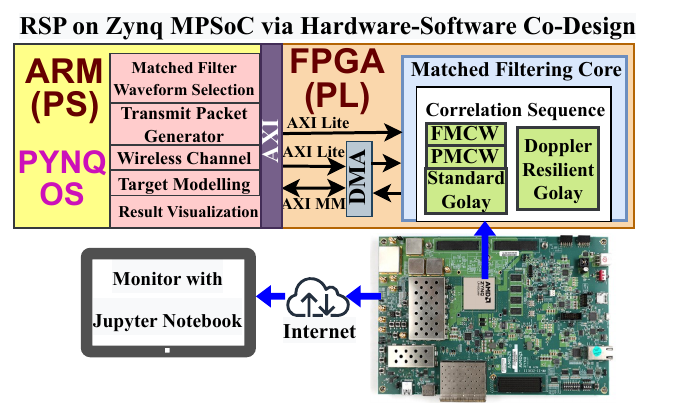}
    \caption{\footnotesize Hardware setup with AMD Xilinx ZCU111 connected to institute network.}
     \label{fig:hardware}
\end{figure}

Mapping RSP algorithms on edge platforms is crucial for validating their feasibility for real-time processing and ease of deployment. In this section, we design and implement the 1D matched filtering algorithm for range estimation of a moving target on Zynq SoC via HSCD and FP analysis. We utilize the Zynq multi-processor SoC (Zynq MPSoC), which includes a quad-core ARM Cortex A53 processor and Ultrascale FPGA from AMD-Xilinx as shown in Fig.~\ref{fig:hardware}. 

\begin{table}[htbp]
\scriptsize
\addtolength{\tabcolsep}{-7pt}
    \centering
    \caption{\footnotesize Resource utilization, power consumption, and acceleration factor comparison of RSP for different PL architectures}
\begin{tabular}{|c|cccc|c|c|}
\hline
\multirow{2}{*}{\begin{tabular}[c]{@{}c@{}}\textbf{Matched}\\ \textbf{Filtering}\end{tabular}} &
  \multicolumn{4}{c|}{\textbf{Resource Utilization}} &
  \multirow{2}{*}{\begin{tabular}[c]{@{}c@{}}\textbf{Dynamic Power} \\ \textbf{Consumption (W)}\end{tabular}} &
  \multirow{2}{*}{\begin{tabular}[c]{@{}c@{}}\textbf{AF} \\ \textbf{w.r.t PS}\end{tabular}} \\ \cline{2-5}
     & \multicolumn{1}{c|}{\textbf{BRAM}} & \multicolumn{1}{c|}{\textbf{DSP}} & \multicolumn{1}{c|}{\textbf{LUT}}   & \textbf{FF}    &       &   \\ \hline
\textbf{SPFL} & \multicolumn{1}{c|}{250}  & \multicolumn{1}{c|}{78}  & \multicolumn{1}{c|}{22024} & 30433 & 3.454 & 9 \\ \hline
\textbf{FP\textless{}\textbf{$24,1$}\textgreater{}} &
  \multicolumn{1}{c|}{\begin{tabular}[c]{@{}c@{}}172\\ \textbf{(-31.2\%)}\end{tabular}} &
  \multicolumn{1}{c|}{\begin{tabular}[c]{@{}c@{}}54\\ \textbf{(-30.7\%)}\end{tabular}} &
  \multicolumn{1}{c|}{\begin{tabular}[c]{@{}c@{}}16759\\ \textbf{(-24.7\%)}\end{tabular}} &
  \begin{tabular}[c]{@{}c@{}}18798\\ \textbf{(-38.2\%)}\end{tabular} &
  \begin{tabular}[c]{@{}c@{}}3.146\\ \textbf{(-8.9\%)}\end{tabular} &
  \begin{tabular}[c]{@{}c@{}}10.6\\ \textbf{(+17.7\%)}\end{tabular} \\ \hline
\end{tabular}
  \label{table:hw_cmpx}
\end{table}
We use HSCD to partition the various tasks between the processing system (PS) or ARM processor and programmable logic (PL) or FPGA on the MPSoC as shown in Fig.~\ref{fig:hardware}. Initially, radar signal modeling, target modeling, and RSP were implemented entirely in PS with double-precision floating point (DPFL). Next, we offload the matched filtering block to the FPGA with single precision floating point (SPFL) architecture. This provides a significant speed-up in execution time as compared to PS but with increased hardware complexity. The matched filtering core in PL stores the different correlation sequences and is configured by PS during runtime to switch between different ISAC waveforms. We explore the fixed-point (FP) architecture to further reduce resource utilization, power consumption, and latency while maintaining performance identical to that of the SPFL architecture. 

\begin{figure}[htbp]
   \centering
    \includegraphics[scale = 0.3]{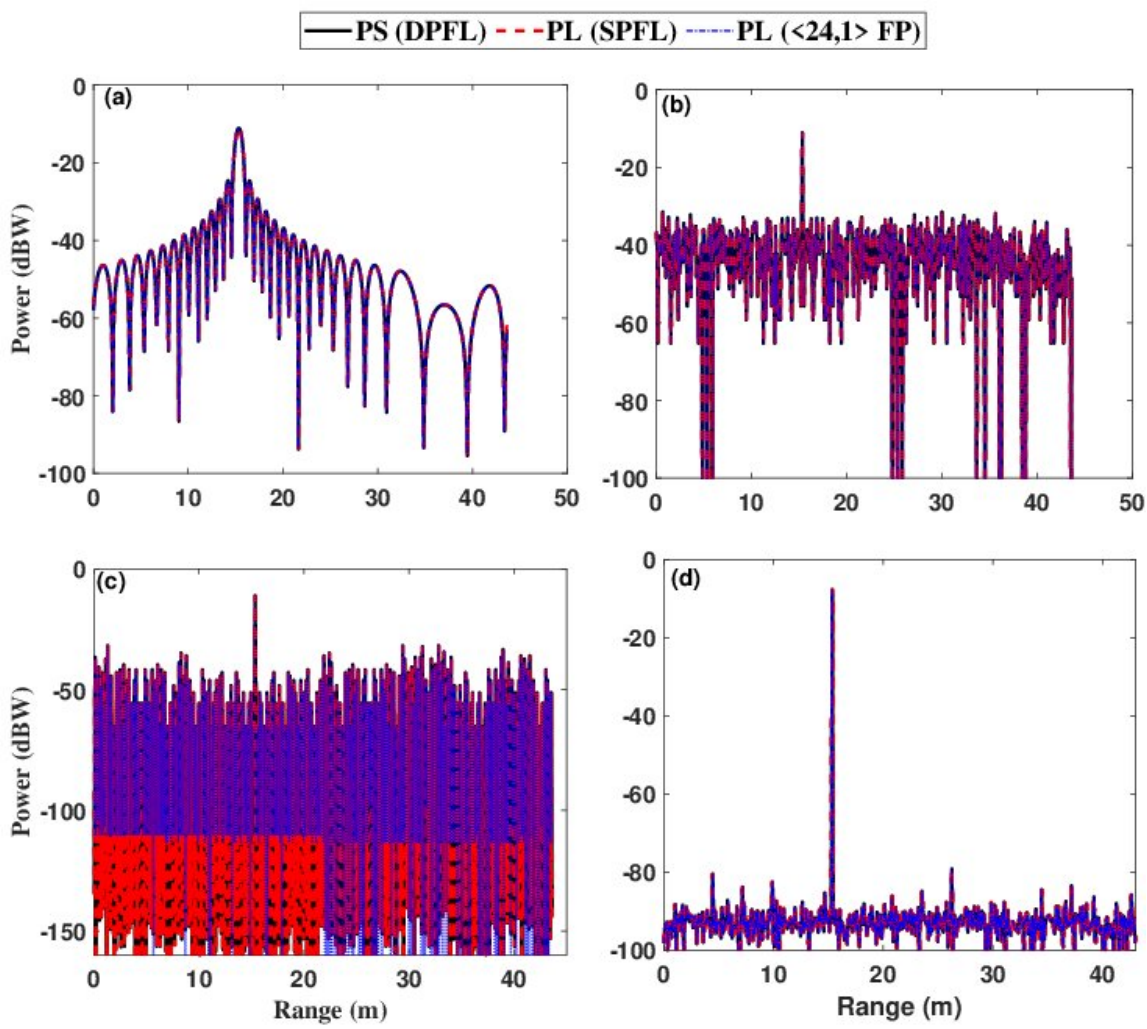}
    \caption{Matched filter outputs for (a) FMCW, (b) PMCW, (c) standard 802.11ad, Golay sequence and (d) Doppler-resilient 802.11ad radars Golay sequence waveforms.}
\label{fig:MF_plot}   
\end{figure}

We consider an isotropic single point scatterer, initially located at Cartesian coordinates of [12,9,0], moving with a velocity of $2$ m/s. Fig.\ref{fig:MF_plot} shows the matched filtered response obtained after range estimation for different waveforms. We observe that the FMCW waveform in Fig.\ref{fig:MF_plot}(a) exhibits poor auto-correlation properties with high sidelobes, resulting in a peak-to-sidelobe ratio (PSLR) of 8dB. The PMCW waveform in Fig.\ref{fig:MF_plot}(b) shows a slight reduction in the side lobe level with PSLR of 13dB. Further, the standard Golay 802.11ad in Fig.\ref{fig:MF_plot}(c) shows a good degree of Doppler resilience for a moving target, leading to 43dB PSLR. Lastly, Fig.\ref{fig:MF_plot}(d) shows that Doppler-resilient Golay 802.11ad does not suffer from significant deterioration in the PSLR in the presence of moving targets. In all of the above cases, we observe that the FP architecture with a selected word length of 24 bits offers an identical range estimation performance compared to the SPFL architecture. Table~\ref{table:hw_cmpx} analyzes the RSP hardware complexity of different PL architectures. It can be seen that the FP \textless{}$24,1$\textgreater{} architecture offers substantial reductions in resource utilization (up to 30\%) and power consumption (up to 9\%) compared to SPFL and an acceleration of up to 10.6 times compared to the PS implementation, without degradation of RSP performance.

\section{Conclusion}
\label{sec:Conclusion}
In this work, we demonstrate the performance of different waveforms that can be used for the localization of the target in ISAC systems for mmW communications. The simulation and the hardware results demonstrate that the  PSLR is highest for the FMCW waveform for both single-point and extended targets. The performance of PMCW is slightly better than that of FMCW. The IEEE 802.11ad waveform shows the perfect auto-correlation property with zero sidelobes. However, the performance of the standard Golay sequence of 802.11ad protocol slightly degrades for a moving target, which is further improved through the Doppler-resilient Golay sequence with zero sidelobes even for a moving target. Further, the FP hardware implementation of RSP shows significant acceleration over DPFL implementation.

\bibliographystyle{IEEEtran}
\bibliography{references}

\end{document}